\documentclass[twocolumn]{aastex631}

\shorttitle{NIRISS determined EWs of rest-optical emission lines in $1 < \rm{z} < 3.4$ SFGs}
\shortauthors{Boyett et al.}
\usepackage{hyperref}  
\hypersetup{colorlinks=true,linkcolor=[rgb]{1.,0.2,0.2},citecolor=[rgb]{0.1,0.4,1.},filecolor=[rgb]{0.7,0.2,0.2},urlcolor=[rgb]{0.7,0.2,0.2}}

\usepackage{color}
\definecolor{blue}{rgb}{0., 0., 1}

\newcommand {\T}{Table\,}

\newcommand{\oiii}{[\textrm{O}\textsc{iii}]}
\newcommand{\oii}{[\textrm{O}\textsc{ii}]}
\newcommand{\oiiiaur}{[\textrm{O}\textsc{iii}]\ensuremath{\lambda4363}}

\newcommand{\oiiiv}{[\textrm{O}\textsc{iii}]\ensuremath{\lambda5007}}

\newcommand{\ha}{\ifmmode {\rm H}\alpha \else H$\alpha$\fi}
\newcommand{\hb}{\ifmmode {\rm H}\beta \else H$\beta$\fi}
\newcommand{\hd}{\ifmmode {\rm H}\delta \else H$\delta$\fi}
\newcommand{\hg}{\ifmmode {\rm H}\gamma \else H$\gamma$\fi}
\newcommand{\lya}{\ifmmode {\rm Ly}\alpha \else Ly$\alpha$\fi}
\newcommand{\pg}{\ifmmode {\rm P}\gamma \else Pa$\gamma$\fi}
\newcommand{\lyb}{\ifmmode {\rm Ly}\beta \else Ly$\beta$\fi}
\newcommand{\lyg}{\ifmmode {\rm Ly}\gamma \else Ly$\gamma$\fi}

\newcommand{\neiii}{\textrm{Ne}\textsc{iii}]\ensuremath{\lambda3869}]}

\newcommand{\flyc}{\ifmmode  \mathrm{f}_\mathrm{esc}\mathrm{(LyC)} \else $\mathrm{f}_\mathrm{esc}\mathrm{(LyC)}$\fi}

\def\ergs{\ifmmode \mathrm{erg\hspace{1mm}s}^{-1} \else erg s$^{-1}$\fi}

\def\micron{\ifmmode \mu\mathrm{m} \else $\mu$m\fi}
\def\msun{\ifmmode \mathrm{M}_{\odot} \else M$_{\odot}$\fi}
\def\msunyr{\ifmmode \mathrm{M}_{\odot} \hspace{1mm}{\rm yr}^{-1} \else $\mathrm{M}_{\odot}$ yr$^{-1}$\fi}
\def\zsun{\ifmmode Z_{\odot} \else Z$_{\odot}$\fi}
\def\lsun{\ifmmode L_{\odot} \else L$_{\odot}$\fi}
\def\mstar{\ifmmode \mathrm{M}_{\star} \else M$_{\star}$\fi}

\newcommand{\hst}{\textit{HST}}
\newcommand{\jwst}{\textit{JWST}}

\begin{document}

\title{Early results from GLASS-JWST. VI: Extreme rest-optical equivalent widths detected in NIRISS Wide Field Slitless Spectroscopy \footnote{Based on observations collected with JWST under the ERS program ID 1324 (PI T. Treu)}}

\correspondingauthor{Kristan Boyett}
\email{kit.boyett@unimelb.edu.au}

%%% Lead authors and main contributors

\author[0000-0003-4109-304X]{K.~Boyett}
\affiliation{School of Physics, University of Melbourne, Parkville 3010, VIC, Australia}
\affiliation{ARC Centre of Excellence for All Sky Astrophysics in 3 Dimensions (ASTRO 3D), Australia}

\author[0000-0002-9572-7813]{S.~Mascia}
\affiliation{INAF Osservatorio Astronomico di Roma, Via Frascati 33, 00078 Monteporzio Catone, Rome, Italy}

\author[0000-0001-8940-6768 ]{L. Pentericci}
\affiliation{INAF Osservatorio Astronomico di Roma, Via Frascati 33, 00078 Monteporzio Catone, Rome, Italy}

\author[0000-0003-4570-3159]{N. Leethochawalit}
\affiliation{School of Physics, University of Melbourne, Parkville 3010, VIC, Australia}
\affiliation{ARC Centre of Excellence for All Sky Astrophysics in 3 Dimensions (ASTRO 3D), Australia}
\affiliation{National Astronomical Research Institute of Thailand (NARIT), Mae Rim, Chiang Mai, 50180, Thailand}

\author[0000-0001-9391-305X]{M. Trenti}
\affiliation{School of Physics, University of Melbourne, Parkville 3010, VIC, Australia}
\affiliation{ARC Centre of Excellence for All Sky Astrophysics in 3 Dimensions (ASTRO 3D), Australia}

% NIRISS builders (see GLASS publication policy doc)
\author[0000-0003-2680-005X]{G. Brammer}
\affiliation{Cosmic Dawn Center (DAWN), Denmark}
\affiliation{Niels Bohr Institute, University of Copenhagen, Jagtvej 128, DK-2200 Copenhagen N, Denmark}

\author[0000-0002-4140-1367]{G. Roberts-Borsani}
\affiliation{Department of Physics and Astronomy, University of California, Los Angeles, 430 Portola Plaza, Los Angeles, CA 90095, USA}

\author[0000-0002-6338-7295]{V. Strait}
\affiliation{Cosmic Dawn Center (DAWN), Denmark}
\affiliation{Niels Bohr Institute, University of Copenhagen, Jagtvej 128, DK-2200 Copenhagen N, Denmark}

\author[0000-0002-8460-0390]{T. Treu}
\affiliation{Department of Physics and Astronomy, University of California, Los Angeles, 430 Portola Plaza, Los Angeles, CA 90095, USA}

\author[0000-0001-5984-0395]{M. Bradac}
\affiliation{University of Ljubljana, Department of Mathematics and Physics, Jadranska ulica 19, SI-1000 Ljubljana, Slovenia}
\affiliation{Department of Physics and Astronomy, University of California Davis, 1 Shields Avenue, Davis, CA 95616, USA}

\author[0000-0002-3254-9044]{K. Glazebrook}
\affiliation{Centre for Astrophysics and Supercomputing, Swinburne
University of Technology, P.O. Box 218, Hawthorn, VIC 3122, Australia}

%%% Additional contributors in alphabetical order

\author[0000-0003-3108-9039]{A.~Acebron}
\affiliation{Dipartimento di Fisica, Università degli Studi di Milano, Via Celoria 16, I-20133 Milano, Italy}
\affiliation{INAF - IASF Milano, via A. Corti 12, I-20133 Milano, Italy}

\author[0000-0003-1383-9414]{P.~Bergamini}
\affiliation{Dipartimento di Fisica, Università degli Studi di Milano, Via Celoria 16, I-20133 Milano, Italy}
\affiliation{INAF - OAS, Osservatorio di Astrofisica e Scienza dello Spazio di Bologna, via Gobetti 93/3, I-40129 Bologna, Italy}

\author[0000-0003-2536-1614]{A. Calabr\`o}
\affiliation{INAF Osservatorio Astronomico di Roma, Via Frascati 33, 00078 Monteporzio Catone, Rome, Italy}

\author[0000-0001-9875-8263]{M.~Castellano}
\affiliation{INAF Osservatorio Astronomico di Roma, Via Frascati 33, 00078 Monteporzio Catone, Rome, Italy}

\author[0000-0003-3820-2823]{A. Fontana}
\affiliation{INAF Osservatorio Astronomico di Roma, Via Frascati 33, 00078 Monteporzio Catone, Rome, Italy}

\author[0000-0002-5926-7143]{C. Grillo}
\affiliation{Dipartimento di Fisica, Università degli Studi di Milano, Via Celoria 16, I-20133 Milano, Italy}
\affiliation{INAF - IASF Milano, via A. Corti 12, I-20133 Milano, Italy}

\author[0000-0002-6586-4446 ]{A. Henry}
\affiliation{Space Telescope Science Institute, 3700 San Martin
  Drive, Baltimore, MD, 21218, USA}
\affiliation{Center for Astrophysical Sciences, Department of Physics \& Astronomy, Johns Hopkins University, Baltimore, MD 21218, USA}

\author[0000-0001-5860-3419]{T. Jones}
\affiliation{Department of Physics and Astronomy, University of California Davis, 1 Shields Avenue, Davis, CA 95616, USA}

\author[0000-0001-9002-3502]{D. Marchesini}
\affiliation{Department of Physics and Astronomy, Tufts University, 574 Boston Ave., Medford, MA 02155, USA}

\author[0000-0002-3407-1785]{C. Mason}
\affiliation{Cosmic Dawn Center (DAWN), Denmark}
\affiliation{Niels Bohr Institute, University of Copenhagen, Jagtvej 128, DK-2200 Copenhagen N, Denmark}

\author[0000-0001-9261-7849]{A. Mercurio}
\affiliation{INAF -- Osservatorio Astronomico di Capodimonte, Via Moiariello 16, I-80131 Napoli, Italy}

\author[0000-0002-8512-1404]{T. Morishita}
\affil{Infrared Processing and Analysis Center, Caltech, 1200 E. California Blvd., Pasadena, CA 91125, USA}

\author[0000-0003-2804-0648 ]{T. Nanayakkara}
\affiliation{Centre for Astrophysics and Supercomputing, Swinburne
University of Technology, P.O. Box 218, Hawthorn, VIC 3122, Australia}

\author[0000-0002-6813-0632]{P.~Rosati}
\affiliation{Dipartimento di Fisica e Scienze della Terra, Università degli Studi di Ferrara, Via Saragat 1, I-44122 Ferrara, Italy}

\author[0000-0002-9136-8876]{C.~Scarlata}\affiliation{
School of Physics and Astronomy, University of Minnesota, Minneapolis, MN, 55455, USA}

\author[0000-0002-5057-135X]{E.~Vanzella}
\affiliation{INAF - OAS, Osservatorio di Astrofisica e Scienza dello Spazio di Bologna, via Gobetti 93/3, I-40129 Bologna, Italy}

\author[0000-0003-0980-1499]{B. Vulcani}
\affiliation{INAF Osservatorio Astronomico di Padova, vicolo dell'Osservatorio 5, 35122 Padova, Italy}

\author[0000-0002-9373-3865]{X. Wang}
\affil{Infrared Processing and Analysis Center, Caltech, 1200 E. California Blvd., Pasadena, CA 91125, USA}

\author[0000-0002-4201-7367]{C. Willott}
\affiliation{NRC Herzberg, 5071 West Saanich Rd, Victoria, BC V9E 2E7, Canada}



\begin{abstract}
Wide Field Slitless Spectroscopy (WFSS) provides a powerful tool for detecting strong line emission in star forming galaxies (SFGs) without the need for target pre-selection.
As part of the GLASS-JWST-ERS program, we leverage the near-infrared wavelength capabilities of NIRISS ($1-2.2\mu$m) to observe rest-optical emission lines out to $z\sim 3.4$, to a depth and with a spatial resolution higher than ever before (\ha\ to $\rm{z}<2.4$; \oiii+\hb\ to $\rm{z}<3.4$). In this letter we constrain the rest-frame \oiiiv\, equivalent width (EW) distribution for a sample of 76 $1<\rm{z}<3.4$ SFGs in the Abell 2744 Hubble Frontier Field and determine an abundance fraction of extreme emission line galaxies with EW$>750$\AA\ in our sample to be $12\%$. We determine a strong correlation between the measured \hb\, and \oiiiv\, EWs, supporting that the high \oiiiv\, EW objects require massive stars in young stellar populations to generate the high energy photons needed to doubly ionise oxygen. We extracted spectra for objects up to 2 mag fainter in the near-infrared than previous WFSS studies with the Hubble Space Telescope. Thus, this work clearly highlights the potential of JWST/NIRISS to provide high quality WFSS datasets in crowded cluster environments. 

\end{abstract}


\section{Introduction} \label{sec:intro}

The study of star forming galaxies (SFGs) has benefited from the development of \hst/WFC3 Wide Field Slitless Spectroscopy (WFSS), where the low background of space-based observatories provides sensitivity to the dispersed light of sources through a low-spectral resolution grism ($\rm{R}=\frac{\lambda}{\Delta \lambda} \sim100-200$). The advantage compared to fixed-slit or multi-object spectroscopy is the ability to obtain spectra for every object in the field of view (FOV) without the need for pre-selection and without being encumbered by slit-losses. Substantial work using WFSS on \hst/WFC3 over the past decade, especially from survey programs such as GLASS \citep{schmidt14, Treu15}, WISPS \citep{Atek10,Bagley2020} and 3D-HST \citep{Momcheva-2016}, has seen the detection of line emission for large galaxy samples over a wide redshift range. With the advent of the James Webb Space Telescope (\jwst) and NIRISS \citep{Doyon12, Willott22}, the observable spectral range available to WFSS now extends further into the near-infrared. 
The additional wavelength range now allows key emission lines to be traced to higher redshifts. The \jwst\, GLASS-ERS program (ERS
1324, PI Treu; \citealt{TreuGlass22}) is one of the first and deepest extragalactic data sets of the Early Release Science (ERS) strategy, and as part of this program we obtain NIRISS/WFSS observations of the Hubble Frontier Field Abell 2744 \citep[A2744,][]{Lotz_2017HFF}. 

Through WFSS, it is possible to examine the
equivalent widths (EW) of rest-optical emission lines, which can be used to understand the conditions of the interstellar medium (ISM) and the nature of the underlying stellar population. WFSS is well leveraged to identify galaxies with strong line emission, including those with weak stellar continua which may be too faint to enter broadband-magnitude selections.
The combination of strong nebular line emission and weak rest-optical stellar continuum is characteristic of galaxies going through an upturn/burst of star formation with a young stellar population. The UV-emission from the massive short-lived O/B-type stars powers the strong nebular lines and, without a dominant older stellar population providing a strong stellar continuum in the rest-optical, these galaxies exhibit large EWs (hereafter Extreme Emission Line Galaxies or EELGs). 

In this letter, we use the GLASS-ERS \jwst/NIRISS observations of the A2744 to study the rest-optical emission line properties of SFGs at $1 < \rm{z} < 3.4$ (where \oiii\, is observable). The A2744 region has been well studied and large samples of galaxies with spectroscopic redshifts at our desired epoch have been previously identified.   Therefore, these objects are the focus of this early work.
In the future we will use the same dataset to also search for emission-line galaxies without previous redshift determination.
In Section \ref{sec:data}, we briefly introduce the \jwst/NIRISS observations.
In Section \ref{sec:indiv}, we use NIRISS/WFSS to measure the rest-optical line emission (\oii, \hb, \oiii\, and \ha)
from which we determine the equivalent widths of these SFGs in Section \ref{sec:method}.
Finally, we discuss the  most extreme line emitters, EELGs, in Section \ref{sec:discuss}. 

We note that since the NIRISS FOV is centered on the Abell 2744 cluster, all objects in our sample will be subject to some degree of magnification due to gravitational lensing; however, in this study we only consider relative quantities for each target (EWs). Therefore our results are independent of magnification, yet gravitational lensing allows us to study fainter galaxies at higher spatial resolution. We refer the reader to \citet[][paper I of this series]{Roberts-Borsani2022} and \citet[][paper VII of this series]{Vanzella2022} for GLASS-ERS studies which provide detailed discussion and utilisation of gravitational lensing. 
In this study we only measure integrated line properties of galaxies and we refer the reader to \citet[][paper IV of this series]{Wang2022} for a GLASS-ERS study of spatially resolved emission line properties.
Where applicable, we use \textit{H}$_{0}=$70 km/s/Mpc, $\Omega_{m}=$0.3, and $\Omega_{\wedge}=$0.7. All magnitudes are in the AB system \citep{Oke83}.

\section{JWST/NIRISS imaging and spectroscopy of HFF~A2744} \label{sec:data}

As part of the GLASS-ERS strategy, we use the WFSS mode of the \jwst/NIRISS instrument to conduct 1.0–2.2$\mu$m low-resolution spectroscopy (average spectral resolution of R $\sim 150$) over the A2744 region. The NIRISS component of the GLASS-ERS survey utilises a single pointing centered on the core of the Abell 2744 cluster\footnote{cluster coordinates (J2000) R.A. 00:14:23.4, DEC. $-30:23:26$},
giving a full area of 2.2$^\prime \times 2.2^\prime$. The WFSS observations and associated direct-imaging are obtained using the F115W, F150W and F200W blocking filters. 
For the full details of the GLASS-ERS data acquisition, observing strategy and data reduction we refer the reader to \citet{TreuGlass22} and \citet{Roberts-Borsani2022}. 

At this early stage of the mission, there is ongoing development of \jwst\, calibration files and pipelines, which means the quality of the extracted data will likely improve over time. The results in this paper are based on a NIRISS data reduction that utilised the following pedigree of data products: \texttt{CAL\_VER}=1.5.2, \texttt{CRDS\_VER}=11.13.1 and \texttt{CRDS\_CTX}=jwst\_0881.pmap. The EW measurements we present in this letter are wavelength-independent relative-quantities (ratio of line flux to continuum flux density at the same wavelength) and therefore should be relatively unaffected by future changes to the calibration files. 

\section{NIRISS detection of Rest-Optical emission lines in individual sources} \label{sec:indiv}
In this letter, we will focus on NIRISS detections of rest-optical emission lines in a sample of galaxies that have existing spectroscopic redshifts measurements (based on the detection of rest-UV/optical lines) in the range $1 < \rm{z} < 3.4$, where the \oiii\, doublet falls within the NIRISS wavelength coverage. This robust sample is collated from several ancillary studies of the A2744 region, utilising both ground-based rest-UV MUSE spectroscopy \citep[][86 objects]{Richard21,Bergamini2022}, and space-based rest-optical \hst/WFC3 WFSS from \citet[][15 objects]{Treu15}.
There are 101 spectroscopic candidates within our FOV in our desired redshift range, although we note that 5 may be less robust due to quality flags that mark their redshift as \lq\lq uncertain\rq\rq\ \citep{Bergamini2022} or \lq\lq tentative/possible\rq\rq\ \citep{Treu15, Richard21}.

\begin{figure*}
    \centering
    \includegraphics[width=\textwidth]{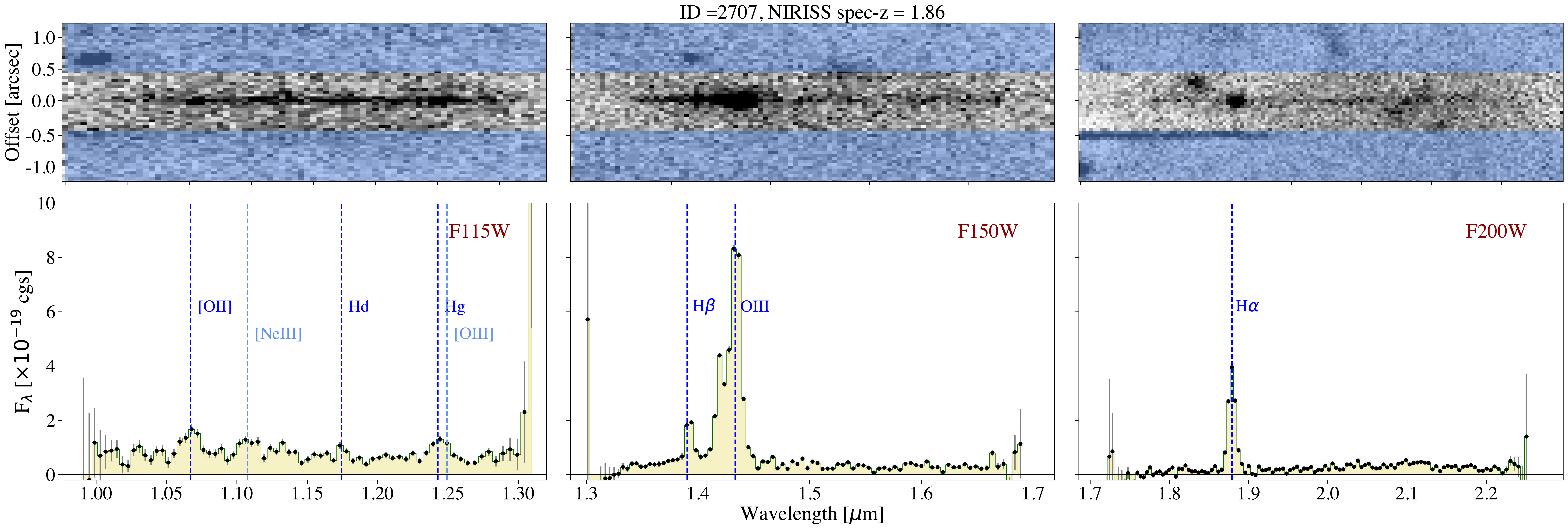}
    \includegraphics[width=\textwidth]{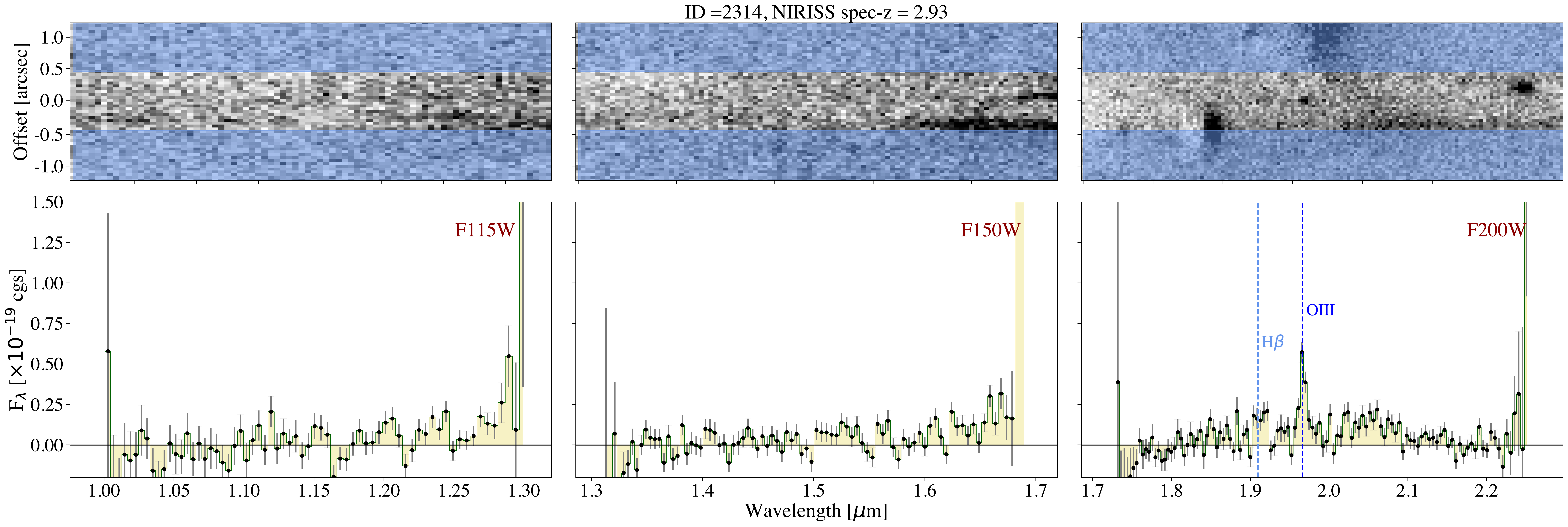}
    \includegraphics[width=\textwidth]{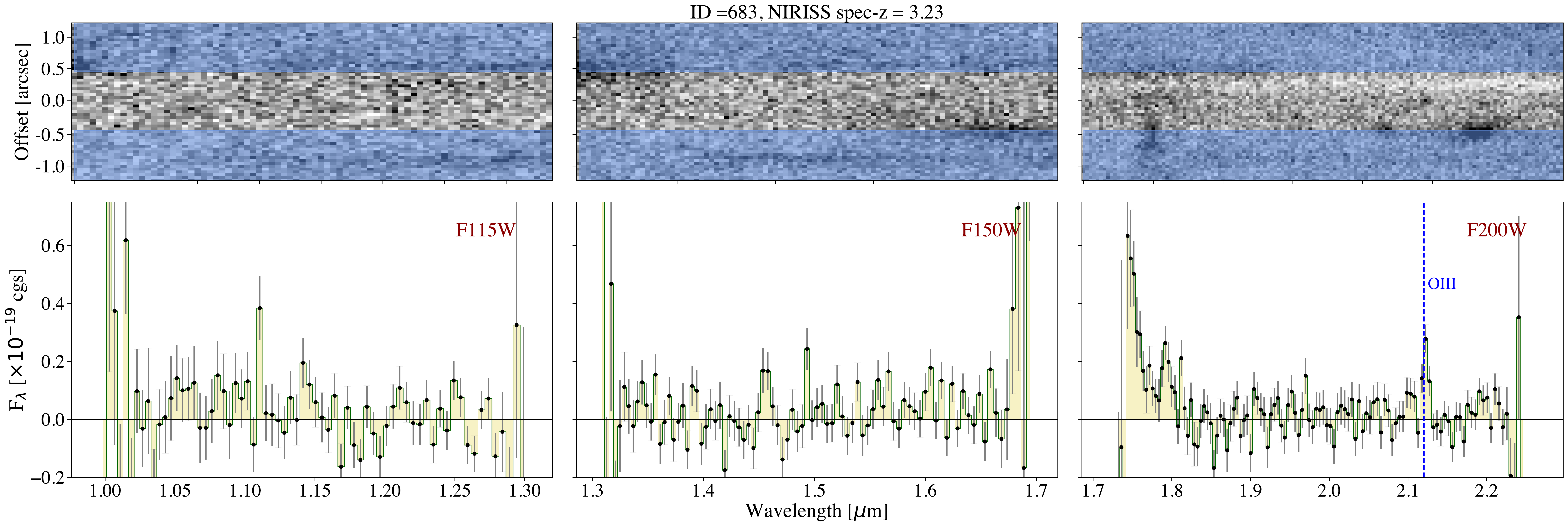}
    \caption{NIRISS/WFSS \texttt{Grizli}-extracted 2D $\&$ 1D spectra for three example galaxies in our sample with large \oiiiv\, EW (these measurements described in Section \ref{sec:indiv}), to first demonstrate the typical high quality spectra and then to show how far we can push the NIRISS data. In the 1D panels we mark the location of $\geq5\sigma$ (dark blue) and $\geq3\sigma$ (light blue) detected emission lines.
    Top: a galaxy with strong detections in each rest-optical emission line; ID2707 ($\rm{z}=1.82$) with a rest-frame \oiiiv EW = $1003\pm37$\AA. The quality of these NIRISS spectra allows the detection of weaker rest-optical lines including \neiii, \hd\, and the blended \hg+\oiiiaur.
    Middle: the galaxy in our full sample with the largest EW; ID2314 ($\rm{z}=2.93$) with a rest-frame \oiiiv\, EW = $1901\pm787$\AA. Here the EW signal-to-noise is $<5$, although we note both the line and broad-band continuum are detected above this threshold. 
    Bottom: the faintest galaxy (F150W total mag = $29.1\pm0.2$) in our full sample with a \oiiiv\, EW measurement; ID683 ($\rm{z}=3.23$) with a rest-frame \oiiiv\, EW = $446\pm109$\AA.
    }
    \label{fig:spectra}
\end{figure*}

Our sample represents only a subset of all the emission line galaxies observed with NIRISS and the selection of objects with existing spectroscopic redshifts, where previous redshift determination was reliant on line detection, may bias our sample against weak line emitters or heavily absorbed objects (which would be too reddened and faint to exhibit strong rest-UV emission lines). For example, there may be a population of \ha\, emitters with no UV counterpart that is missed by this pre-selection.
Consideration of the selection effects and an analysis with a full statistical sample is beyond the scope of this letter and will be performed in the future. Instead, here we focus on the robust sample as a demonstration of the capabilities of \jwst/NIRISS.

We utilise the Grism Redshift and Line Analysis tool \citep[\texttt{Grizli\footnote{\url{https://github.com/gbrammer/grizli}}};][]{brammer21} to extract and model the Grism spectra associated with each of our targets (see \citealt{Roberts-Borsani2022} for further details).
For 16 objects (all selected from rest-UV spectroscopy), we found no coordinate matches within 1$\arcsec$ in the segmentation map of the composite NIRISS F115W+F150W+F200W direct-image and we remove these from our analysis. 

From the 85 matched objects with extracted spectra, we perform the \texttt{Grizli} modelling twice, first with a broad ($0.5 < \rm{z} < 5.0$) redshift fitting range and again with a narrow ($\Delta \rm{z}\pm0.1$) redshift fitting range centred on either the resultant \texttt{Grizli} redshift or the existing spectroscopic redshift if no lines are detected with NIRISS. The first run is used to identify cases when alternative redshift solutions may be present due to new NIRISS line detections, and the second is to aid the identification of weaker lines by reducing the free parameters in the modelling once a confident redshift solution is found. 

Within the 2.2$\arcmin \times$2.2$\arcmin$ area, 76 of the 85 galaxies have a NIRISS ($>5\sigma$) line detection. Of these, 54 have multiple NIRISS line detections. 
The uncertainty in the line flux reported by \texttt{Grizli} reflects instrumental effects and modelling procedure.
We note that uncertainties on faint lines also reflect the presence of additional strong lines in the spectrum which can corroborate the redshift solution.
Quoting a line flux limit for the entire survey is not feasible, as the flux limit  depends on the level of contamination from nearby sources. Accordingly, we provide the observed median 5$\sigma$ line flux limit, with associated 16$^{\rm{th}}$ and 84$^{\rm{th}}$ percentiles. We measure line flux limits of  $1.5^{+1.9}_{-0.5} \times 10^{-17}$, $1.0^{+1.9}_{-0.4} \times 10^{-17}$ and $7.6^{+9.1}_{-3.1} \times 10^{-18} $~erg\,s$^{-1}$\,cm$^{-2}$ for the F115W, F150W and F200W filters respectively using the stated calibration files (these are subject to change in line with calibration file updates). 

From the initial run, 23 sources with detected line emission have a primary redshift solution that disagrees with the previous measurement. Upon inspection, we categorise these into three groups: we believe 2 objects\footnote{NIRISS ID:2808 and 1911. These NIRISS ID's refer to the provided machine readable table which contains the rest-optical properties and coordinates.} have previously mis-identified redshifts based on the observed wavelengths of line emission detected with NIRISS and in the previous spectroscopy; 7 objects have a close-probability secondary redshift solution from \texttt{Grizli} that agrees with the existing value and is supported by additional line detections in previous spectroscopy, and in the second run we utilise their secondary \texttt{Grizli} solution; finally 14 objects have solutions based on weak ($<10\sigma$) detections of only one (7) or two (7) lines in NIRISS which can allow the redshift solution to remain ambiguous, and for these we rely on the existing spectroscopic redshift for the second run.
The 9 galaxies with no NIRISS line-detection either have weak rest-optical line emission, lie at alternative redshifts where no lines are present in our wavelength window, or have considerable contamination from the dispersed light from neighbouring sources such that their lines are no-longer detectable. We note that of the 5 galaxies with less robust redshift quality flags, 3 have their redshift confirmed, 1 is identified to lie at an alternative redshift and 1 had no line detections. This indicates that some of the targets without NIRISS-detections may be due to the quality of the input spectroscopic redshifts. 
Finally, we inspect one object (NIRISS ID:1710) which had two different spectroscopic redshifts based on claimed \oii-line detections from MUSE ($\rm{z}=1.2$) and \hst/WFC3 WFSS ($\rm{z}=1.9$). From our detection of \ha\, and \oiii\, in NIRISS, we measure $\rm{z}_{NIRISS}=1.16$ and propose that the \hst\, line was actually the mis-identified \oiii\, line. 

\begin{figure*}
    \centering
    \includegraphics[width=\textwidth]{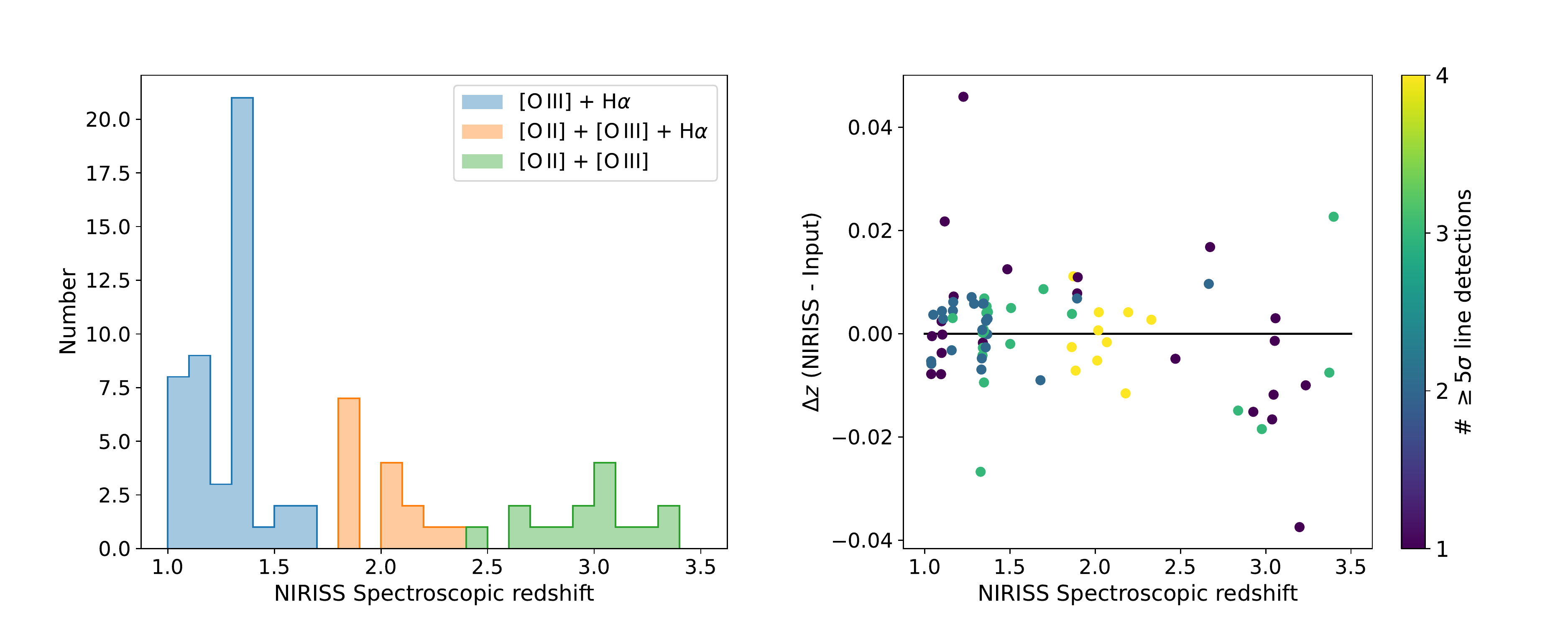}
    \caption{
    Left Panel: The $1<\rm{z}<3.4$ NIRISS spectroscopic redshift distribution of 76 galaxies with NIRISS-detections ($>5\sigma$) of \ha, \oiii, \hb\, or \oii. The distribution is split into three colours based on the set of emission lines which are covered by the observations at different redshifts (e.g., \oii\, is only visible at $\rm{z}>1.7$ and \ha\, at $\rm{z}<2.4$).
    We note an apparent over-density of 21 independent galaxies in the range $1.33<\rm{z}<1.37$.
    Right Panel: The comparison of the NIRISS \texttt{Grizli}-derived and input spectroscopic redshifts, colour coded by the number of $>5\sigma$ NIRISS line detections. Excellent agreement is found in the sample, with the majority of targets having their previous redshifts confirmed through multiple NIRISS line detections. Only two galaxies (NIRISS ID:2808 and 1911, $\Delta \rm{z} > 0.5$) are found to have previously mis-identified redshifts solutions. }
    \label{fig:redshift_dist}
\end{figure*}

In Figure \ref{fig:spectra} we present the 2D and 1D NIRISS spectra for 3 objects that demonstrate the capabilities of \jwst/NIRISS: ID 2314 which had the largest measured \oiiiv\, EW in our sample (determined in the next Section), ID 683 which was the faintest source (F150W $29.1\pm0.2$ mag\footnote{We note a consistent magnitude offset is observed between the NIRISS F150W total-magnitudes and the associated \hst\ F160W total-magnitudes (F160W - F150W = +0.5$\pm$0.1 mag) for the sources from \citet{Bergamini2022}. We apply this offset correction to the quoted magnitude. This discrepancy is likely to be automatically solved once updated NIRISS in-orbit calibration data become available.}) with a NIRISS line detection and ID 2707 which shows significant detections in all rest-optical lines considered in this work. 
The distribution of \texttt{Grizli}-derived spectroscopic redshifts for galaxies with NIRISS line detections is shown in the top panel of Figure \ref{fig:redshift_dist}; redshifts are taken from the second \texttt{Grizli} modelling. 
In the bottom panel the previous spectroscopic redshifts and the \texttt{Grizli}-derived NIRISS redshifts are compared. Excellent agreement is found between the two spectroscopic redshift measurements, with NIRISS line detections confirming 87\%~(74/85) of the sample and only identifying 2 galaxies to be at alternative redshifts.
We note a small $\sigma_{\rm{NMAD}}$\footnote{$\sigma_{\rm{NMAD}}$ as defined by \citet{Brammer_2008}. $\sigma_{\rm{NMAD}} = 1.48 \times \rm{median}(\left | \Delta \rm{z} - \rm{median} (\Delta \rm{z}) \right | / (1+\rm{z}_{\rm{NIRISS}}) )$, where $\Delta \rm{z} = \rm{z}_{\rm{NIRISS}} - \rm{z}_{\rm{INPUT}}$} $= 0.003$ and only 14\%~ (11/76) of those with line detections show notable deviation in spectroscopic redshift ($\Delta \rm{z} / (1+\rm{z}_{\rm{NIRISS}}) > 5\sigma_{\rm{NMAD}}$). 

In summary, we are able to recover accurate redshifts using NIRISS for nearly the entire sample, but note that visual inspection is required in a significant fraction of cases to clearly identify the emission lines. Nonetheless these are early results and the performance of automated redshift and line flux measurements is likely to improve with updated NIRISS calibration and contamination modelling.

\section{Determination of rest-Optical equivalent widths}
\label{sec:method}
To determine individual EW, we start by measuring the integrated line fluxes for emission lines detected at $\geq5\sigma$ in NIRISS/WFSS and place upper limits on those below this threshold. In this analysis, we only consider the 76 galaxies that have at least one NIRISS line detection and emission lines within the NIRISS wavelength range.
For the objects in the field with NIRISS-detected line emission (66/76 have \oiii\ detections), we determine the EW as the ratio of the line flux to the continuum flux density (at the wavelength of the line) where both quantities are \lq\lq total\rq\rq\ values. 

We elect to measure the continuum flux density from the broadband imaging rather than directly from the WFSS spectra. This is because the continuum sensitivity in the spectra 
and the contamination due to overlap with other spectra can lead to large uncertainties on the measured EW for objects with faint continua. 

Total line fluxes and their uncertainty are obtained from the \texttt{Grizli} extraction pipeline, along with broad-band photometry for each source in the F115W, F150W, and F200W direct images. Photometric fluxes are obtained from a 0.36$\arcsec$-diameter aperture with a total-aperture correction applied. Using aperture-corrected continuum and emission line fluxes results in galaxy-integrated EW values.

To validate our approach of combining line fluxes from the spectra with continuum from photometry, we used the bright-end of our sample, where EW measurements from the spectra are robust, and we found EWs in good agreement with those determined using broad-band imaging, i.e. the measurements agree within $\sim$ 0.1 dex ($1\sigma$) with a minor systematic bias of 0.06 dex for 16 objects with $21<m_{AB}<23$.

All galaxies for which we identify emission lines are well detected ($\geq5\sigma$) in the associated imaging, and are thus suitable for equivalent width measurements. The faintest source with a NIRISS detection has a F150W $29.1\pm0.2$ magnitude (compared to a sample median F150W magnitude of 24.57), consistent with the estimated $5\sigma$ magnitude limit for the pre-imaging \citep{TreuGlass22}.
We assume a flat in $f_\nu$ continuum slope and subtract the flux contribution from any detected emission lines contributing to that filter. 
EWs are then taken as the ratio of the line flux to the continuum flux density, and are brought to the rest-frame using the NIRISS \texttt{Grizli}-derived redshift.
The assumed continuum slope and the contribution of non-detected emission lines does not significantly affect the measured EW distribution (e.g., see \citealt{Boyett22}). 

When an emission line is not detected (10/76 do not have \oiii\ detections), we report a $5\sigma$ upper limit on the line flux using the modelled line uncertainty.
The $5\sigma$ EW upper limit is then taken as the ratio of this flux limit and the continuum flux density at the expected location.
Here, we rely on the spectroscopic redshift derived from other NIRISS-detected lines (\ha, \hb, \oii). 

\begin{figure*}
    \centering
    \centerline{\includegraphics[width=\textwidth]{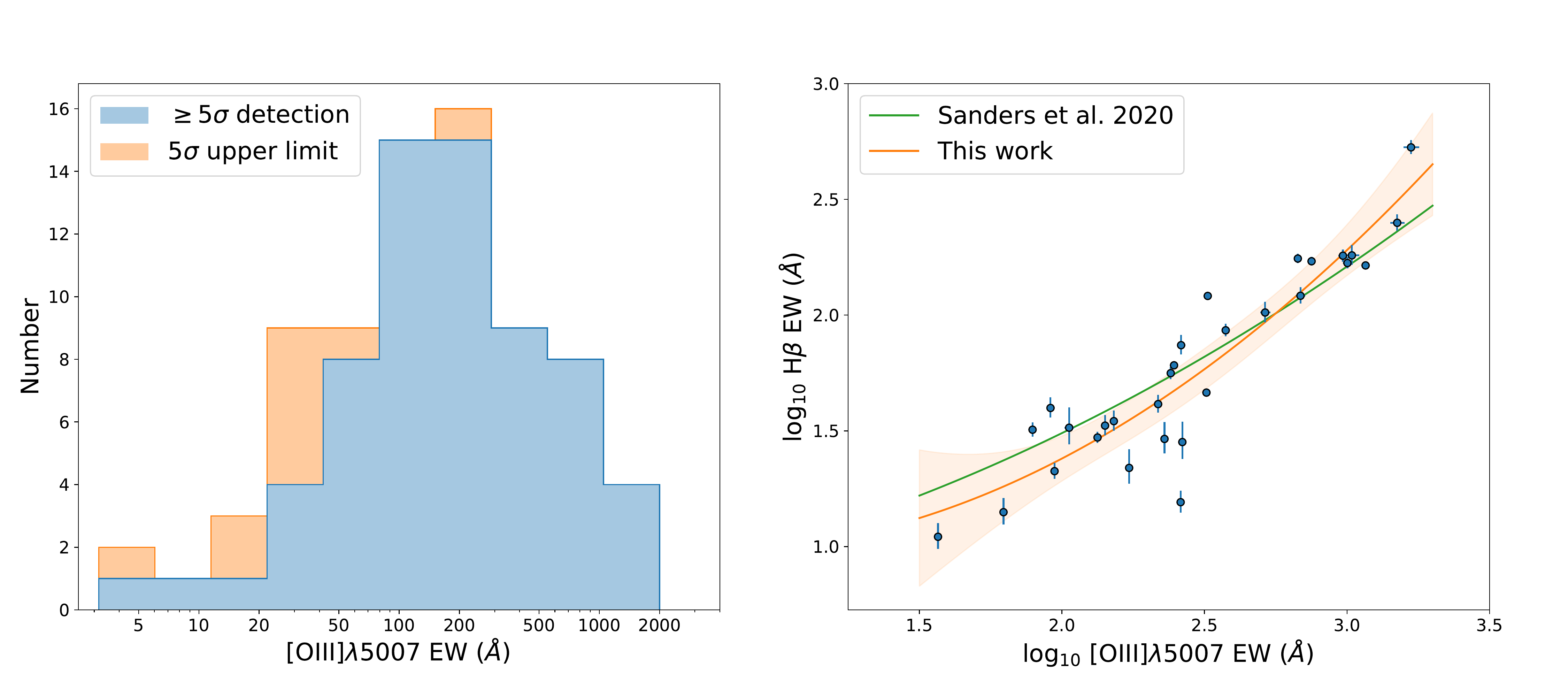}}
    \caption{Rest-optical EW properties of $1 < \rm{z} < 3.4$ SFGs. Left: The \oiiiv\, rest-frame equivalent width distribution of galaxies with NIRISS-detected ($\geq5\sigma$) line emission. Stacked histograms show $\geq5\sigma$ EW measurements (blue) and $5\sigma$ upper limits (orange), divided between 10 bins of equal logarithmic width. 
    Right: The comparison of \hb\, and \oiiiv\, EWs, restricted to objects where both have $\geq5\sigma$ measurements. Overlaid are the $\rm{z}\sim2.3$ \citet{Sanders20} curve (green) and a best-fit second order polynomial with associated $2\sigma$ uncertainty for this data (orange, shaded). }
    \label{fig:EW_dist}
\end{figure*}

We present the distribution of \oiiiv\, EWs in the $1<\rm{z}<3.4$ sample in Figure \ref{fig:EW_dist}, adopting a 1:3 flux ratio between the doublet.
We observe a broad dynamic range of \oiiiv\ EW spanning $\sim$10--2000~\AA, with a clear peak at $\sim$150~\AA. We note that our sample is likely incomplete at low EWs due to our line sensitivity and sample selection; we intend to revisit the completeness in future work. We note 14 objects with \oiii\ EW measurements are at $2.4 < \rm{z} < 3.4$, an epoch previously unobtainable with WFSS on \hst.

We additionally present the comparison of the EWs for \hb\, and \oiiiv\, for galaxies where both have $\geq5\sigma$ measurements. 
We observe a clear positive correlation which is in agreement with the trend found by \citet[][MOSDEF, ground-based]{Sanders20} for $\rm{z}\sim2.3$ SFGs, supporting the accuracy of our measurements. The NIRISS sensitivity provides us higher precision EW measurements and following \citet{Sanders20} we fit a 2$^{\rm{nd}}$ order polynomial and report best fit coefficients: $\log_{10}(\mathrm{EW_{\hb}}) = 0.26x^2 - 0.39x + 1.12$, with $x = \log_{10}(\mathrm{EW_{\oiiiv}})$, and measure a scatter of $0.16$dex.
The \oiii\, and \hb\ EW measurements are provided in Table \ref{tab:summary} for the subset of objects where all rest-optical emission lines are observable and in a machine readable format for the full dataset.

\section{Identification of Extreme Emission Line Galaxies with NIRISS WFSS} \label{sec:discuss}
While there is no strict EW threshold for what is considered an EELG, we will adopt a rest-frame $>750$\AA\,\oiiiv\, EW criteria. This threshold is physically motivated by SFGs above this value being observed to exhibit larger ionising photon production efficiencies than typical SFGs \citep[e.g., ][at z$\,\sim2$]{Mengtao19} and the potential for high escape fractions of ionising photons \citep[e.g., ][]{Nakajima19}. These systems with large EWs are associated with very young stellar populations ($\lesssim10$Myr, e.g., \citealt{Mengtao20}), created during a recent burst of star formation, where the massive O/B-type stars responsible for the production of the high-energy ionising UV photon required to produce doubly ionised oxygen still remain in the population. 

We find that out of the 76 initial targets with NIRISS line detections, 12\% ~(9/76) are considered EELGs according to our threshold.
This value is larger than the measured rate (3.8\% $>750$\AA) at $\rm{z}\sim2$ from \citet[][]{Boyett22}, who utilise \hst/WFC3 WFSS. 
We note that these two values may differ due to the sample selection, given that the quality of the NIRISS/WFSS allows us to extract spectra from sources up to two magnitudes fainter (at $\lambda\sim1.5\mu$m) and \citet[][]{Boyett22} employ a M$_{\rm{UV}}$ restriction. 
Additionally, the requirement for our sample to have a spectroscopic redshift based on previous line detections may bias our sample towards stronger line emitters, although we may also miss extreme line emitters with faint continua due to the requirement of a detection in our direct imaging. Future work will consider the completeness in detail.

Our NIRISS/WFSS sample allows us to study the characteristics of these SFGs. 
\hb\, EW is shown to be a good indicator of the age of a stellar population \citep[e.g.,][]{Flury22, Flury22b} and in this sample we determine a strong relation between the \oiiiv and \hb\, EW.
This supports that very young stars are required in our large-EW SFGs to produce the high energy photons needed for the production of doubly ionised oxygen.

We have provided a first study of the rest-optical properties of EELGs based on NIRISS observations, and to higher redshifts than achievable with \hst/WFC3 WFSS. 
We await upcoming \jwst/NIRISS programs, including, CANUCS [GTO 1208, PI Willott] and PASSAGE [GO 1571, PI Malkan], to provide greater statistical analysis through large area surveys (involving multiple NIRISS pointings).\\

\begin{deluxetable*}{cccccccc}
\tablenum{1}
\tablecaption{Rest-optical emission line EWs for the $1.7<\rm{z}<2.4$ (were all lines are observable) NIRISS-detected sources \label{tab:summary}}
\tablewidth{0pt}
\tablehead{
\colhead{ID} & \colhead{RA} & \colhead{DEC} & \colhead{z} & \colhead{\ha\ EW} & \colhead{\oiii\ EW} & \colhead{\hb\ EW} & \colhead{\oii\ EW}  \\
\colhead{[NIRISS]} & \colhead{[Deg]} & \colhead{[Deg]} & \colhead{[NIRISS]} & \colhead{[\AA]} & \colhead{[\AA]} & \colhead{[\AA]} & \colhead{[\AA]} \\
\colhead{(1)} & \colhead{(2)} & \colhead{(3)} & \colhead{(4,5,6)} & \colhead{(7,8)} & \colhead{(9,10)} & \colhead{(11,12)} & \colhead{(13,14)}}
\startdata
1911 & 3.586765 & -30.390825 & $2.213^{+0.001}_{-0.003}$ & $80.37 \pm 4.77$ & $70.56 \pm 3.54$ & $<8.13 $ & $24.95 \pm 3.78$ \\
2205 & 3.586972 & -30.387021 & $1.864^{+0.001}_{-0.001}$ & $230.41 \pm 12.12$ & $305.21 \pm 8.46$ & $29.16 \pm 4.50$ & $<28.36 $ \\
1208 & 3.576590 & -30.399050 & $1.872^{+0.001}_{-0.001}$ & $249.00 \pm 8.08$ & $500.33 \pm 13.71$ & $85.95 \pm 5.35$ & $45.07 \pm 4.18$ \\
2707 & 3.585922 & -30.380683 & $1.862^{+0.001}_{-0.001}$ & $980.57 \pm 33.70$ & $1337.72 \pm 48.96$ & $167.85 \pm 9.36$ & $106.76 \pm 8.94$ \\
1426 & 3.597289 & -30.396712 & $1.898^{+0.007}_{-0.002}$ & $<123.42 $ & $149.15 \pm 17.34$ & $<62.31 $ & $<68.84 $ \\
951 & 3.583265 & -30.403339 & $1.894^{+0.003}_{-0.002}$ & $212.63 \pm 21.71$ & $109.76 \pm 12.03$ & $<51.88 $ & $81.09 $ \\
950 & 3.582530 & -30.402313 & $1.895^{+0.002}_{-0.001}$ & $<94.77 $ & $211.30 \pm 10.08$ & $<40.78 $ & $<111.51 $ \\
2616 & 3.591120 & -30.381703 & $1.885^{+0.001}_{-0.001}$ & $<287.23 \pm 3.51$ & $428.25 \pm 4.65$ & $46.24 \pm 1.71$ & $41.77 \pm 2.04$ \\
320 & 3.586433 & -30.409363 & $2.021^{+0.001}_{-0.001}$ & $772.55 \pm 19.82$ & $1001.00 \pm 26.51$ & $170.83 \pm 7.23$ & $56.85 \pm 3.89$ \\
1002 & 3.598541 & -30.401792 & $2.012^{+0.001}_{-0.001}$ & $870.69 \pm 25.02$ & $895.90 \pm 26.63$ & $175.48 \pm 8.22$ & $53.87 \pm 4.53$ \\
451 & 3.594058 & -30.407999 & $2.018^{+0.001}_{-0.001}$ & $2631.01 \pm 145.88$ & $2235.78 \pm 140.12$ & $530.46 \pm 36.99$ & $142.72 \pm 9.48$ \\
1352 & 3.604190 & -30.397187 & $2.068^{+0.001}_{-0.001}$ & $163.38 \pm 4.50$ & $105.14 \pm 2.80$ & $31.96 \pm 2.24$ & $34.00 \pm 2.59$ \\
2289 & 3.579076 & -30.385963 & $2.194^{+0.001}_{-0.001}$ & $607.38 \pm 28.01$ & $688.04 \pm 27.85$ & $102.58 \pm 10.39$ & $70.52 \pm 8.02$ \\
1886 & 3.603156 & -30.391074 & $2.178^{+0.001}_{-0.001}$ & $601.62 \pm 24.57$ & $915.38 \pm 35.40$ & $121.17 \pm 9.82$ & $57.73 \pm 5.88$ \\
2139 & 3.580030 & -30.387832 & $2.329^{+0.001}_{-0.001}$ & $218.86 \pm 4.02$ & $347.69 \pm 4.43$ & $15.54 \pm 1.70$ & $97.50 \pm 2.66$  \\
\hline
\enddata
\tablecomments{The columns provided in the machine readable table for the full $1<\rm{z} < 3.4$ sample match those given here. Column (1) lists the IDs of the sources; (2) and (3) show the (J2000) Right Ascension and Declination; (4,5,6) report the \texttt{Grizili} maximum probability function redshift and the 16$^{th}$ and 84$^{th}$ redshift percentiles; the rest-frame equivalent width and uncertainty are given for the \ha, \oiii, \hb\, and \oii\, lines in columns (7-14). When $5\sigma$ EW upper limits are required, the EW\_err is set to -99.}
\end{deluxetable*}

\section{conclusions}

The GLASS-ERS program provides some of the earliest and deepest \jwst\, extragalactic observations, and in this Letter we have used deep NIRISS/WFSS to study $1<\rm{z}<3.4$ SFGs in the A2744 region. From our initial spectroscopic sample of galaxies, we detect line emission in $89\% ~(76/85)$ of targets with extracted spectra. We find good agreement in the spectroscopic redshift measurements from NIRISS and previous ground- and space-based studies, confirming the redshift of 74 objects and with only 2 objects identified using NIRISS to have an alternative redshift to their previous values. We note there is still subtle differences in the spectroscopic redshifts with $14\%$ of the sample ($11/76$ objects) having a change in measurement greater than $\geq 5\sigma_{\rm{NMAD}}$ where $\sigma_{\rm{NMAD}} = 0.003$.
We caution that visual inspection was required in a significant fraction of cases (23/76) to clearly identify the emission lines. This will likely improve with updates to contamination modelling and NIRISS calibrations.

We determine the observed rest-frame \oiiiv\, EW distribution, ranging from $\sim$10--2000~\AA, with a turnover at $\sim$150~\AA.
We identify an EELG sub-sample of $9~ (12\%)$ galaxies with EW $\geq750$\AA, an abundance higher than found in previous studies. This discrepancy likely arises from the inherent incompleteness of our sample due to the pre-selection of spectroscopically identified sources and our line sensitivity, although we note the sensitivity of NIRISS allows our sample to include extracted spectra from galaxies up to 2 magnitudes fainter in the near-infrared and to higher redshift than previous \hst/WFSS studies.
Finally, we find a correlation between the \oiiiv\, and \hb\, EWs, supporting that very young stellar populations are required to produce the high energy photons needed to produce \oiii. 

This study provides a first demonstration of the capabilities of \jwst/NIRISS for studying rest-optical emission lines in high-redshift SFGs. 
We anticipate that future studies of SFG spectra through NIRISS/WFSS will advance greatly over the first cycles of \jwst\, science, as planned wide area surveys including PASSAGE and CANUCS unlock large galaxy samples, allowing statistical analysis of the rare EELG population.

%% Also note that the akcnowlodgment environment does not support long amounts of text. If you have a lot of people and institutions to acknowledge, do not use this command. Instead, create a new \section{Acknowledgments}.
\section{Acknowledgments}
This work is based on observations made with the NASA/ESA/CSA James Webb Space Telescope. The data were obtained from the Mikulski Archive for Space Telescopes at the Space Telescope Science Institute, which is operated by the Association of Universities for Research in Astronomy, Inc., under NASA contract NAS 5-03127 for JWST. These observations are associated with program JWST-ERS-1324. We acknowledge financial support from NASA through grant JWST-ERS-1324.  KG and TN acknowledge support from Australian Research Council Laureate Fellowship FL180100060. This research is supported in part by the Australian Research Council Centre of Excellence for All Sky Astrophysics in 3 Dimensions (ASTRO 3D), through project number CE170100013. We acknowledge financial support through grants PRIN-MIUR 2017WSCC32 and 2020SKSTHZ. MB acknowledges support from the Slovenian national research agency ARRS through grant N1-0238. CM acknowledges support by the VILLUM FONDEN under grant 37459. The Cosmic Dawn Center (DAWN) is funded by the Danish National Research Foundation under grant DNRF140.

\section*{Data Availability}
We provide measured rest-optical emission line properties (Table \ref{tab:summary}) for the full sample in a machine readable format.
Further catalogues and the routines used in this work are available from the authors upon reasonable request. The raw \jwst\, observations are publicly available from STSCI.

\bibliography{ref}{}
\bibliographystyle{aasjournal}



\end{document}